\documentclass{article}

\usepackage[english]{babel}
\usepackage[letterpaper,top=2cm,bottom=2cm,left=3cm,right=3cm,marginparwidth=1.75cm]{geometry}
\usepackage{amsmath}
\usepackage{graphicx}
\usepackage{float}
\usepackage{changepage}
\usepackage{tabularx}
\usepackage{booktabs}
\usepackage{multirow}
\usepackage{array}
\usepackage{subfig}
\usepackage{cite}
\usepackage{physics}
\usepackage{authblk}
\usepackage[colorlinks=true, allcolors=blue]{hyperref}
\usepackage{orcidlink}

\title{\bfseries\Large Robust Atom Interferometry with Super-Gaussian Pulses against Thermal Velocity Spread}

\author[1]{Yujuan Liu~\orcidlink{0009-0009-3456-1272}} 
\author[1]{Ziwen Song~\orcidlink{0009-0009-3321-1131}}      
\author[1, *]{Tingting Lin~\orcidlink{0000-0002-6061-2311}}
\author[2]{Biao Tang~\orcidlink{0009-0006-9047-9060}}
\author[1]{Aoxing Hao~\orcidlink{0009-0006-9627-9430}}

\affil[1]{Jilin University, College of Instrumentation and Electrical Engineering, Changchun, China}
\affil[2]{Division of Precision Measurement Physics, Innovation Academy for Precision Measurement Science and Technology, Chinese Academy of Sciences, Wuhan 430071, China}
\affil[*]{ttlin@jlu.edu.cn}

\date{\fontsize{8}{11}\selectfont (Dated: \today)}
\begin{document}

\maketitle

\begin{abstract}
Laser frequency fluctuation and atomic thermal motion can lead to errors in pulse duration and detuning in cold atom interferometry, thereby reducing measurement stability and fringe contrast. To address this issue, we investigate the use of super-Gaussian pulses, which are characterized by smooth temporal profiles and centralized energy distribution, in the beam-splitting and reflection stages of an atom interferometer. Through numerical simulations, we compare the performance of rectangular, Gaussian, and 2nd- to 10th-order super-Gaussian pulses subject to deviations in pulse duration and detuning. Our results show that both Gaussian and super-Gaussian pulses offer a significant advantage over traditional rectangular pulses, particularly under thermal conditions where velocity spread is prominent. We find that 4th-order pulses achieving up to a 90\% improvement in contrast over rectangular pulses under realistic conditions, and while their peak performance at very low temperatures is comparable to that of Gaussian pulses, they demonstrate enhanced robustness against combined detuning and pulse-length errors. These findings demonstrate that super-Gaussian pulse shaping is an effective method for enhancing the robustness of atom interferometers against errors induced by thermal motion.
\end{abstract}

\centerline{\textbf{Keywords:} atom interferometry, atomic physics, pulse shaping}

\newpage

\section{Introduction}

Cold atom interferometers\cite{berman1997atom, Baudon_1999} represent a state-of-the-art technique in precision metrology, based on the quantum coherent superposition of matter waves. By cooling and coherently manipulating atomic clouds to generate interference effects, they can surpass the standard quantum limit imposed on conventional optical interferometers, offering superior sensitivity and precision\cite{Giovannetti_2004}. Atom interferometry based on matter-wave interference enables ultra-high-precision measurements across a broad range of applications, including linear acceleration\cite{rakholia2014dual, canuel2006six}, rotation\cite{dutta2016continuous,gustavson2000rotation}, and the Earth's gravitational field\cite{louchet2011influence}, as well as its gradient and curvature\cite{mcguirk2002sensitive,rosi2015measurement}. These systems are also employed in the precise determination of fundamental physical constants\cite{rosi2014precision,lamporesi2008determination} and in tests of the weak equivalence principle\cite{asenbaum2020atom}.

In light-pulse atom interferometry, a $\pi/2$-$\pi$-$\pi/2$ laser pulse sequence coherently splits, redirects, and recombines atomic wave packets. Achieving optimal sensitivity—either by enlarging the interferometer area or maintaining quantum coherence—requires that the $\pi/2$ and $\pi$ pulses operate with high fidelity. However, imperfections arising from electromagnetic field fluctuations, laser intensity or frequency fluctuation, atomic velocity spread, and quantum state inhomogeneity lead to pulse area and detuning errors, which degrade the fidelity of quantum state manipulation. While quantum state inhomogeneity relates to an imperfect initial population distribution, the atomic velocity spread, resulting from thermal motion, introduces a distribution of detuning errors across the atomic ensemble. Both of these effects limit the number of coherent control operations and reduce the interference fringe contrast.\cite{wu2019gravity,bidel2013compact}. These imperfections limit the number of coherent control operations that can be performed before decoherence occurs.

To mitigate such effects, composite\cite{van2014interferometry,jager2014optimal} and shaped pulse\cite{daems2013robust,szigeti2012momentum,muller2008atom,luo2016contrast} techniques—originally developed in quantum information processing\cite{peterson2020enhancing,cross2015optimized} and nuclear magnetic resonance (NMR)\cite{vandersypen2004nmr,cummins2003tackling}—have been employed to generate control pulses robust to fluctuations in interaction strength and detuning. These methods are also applicable to cold atom interferometers. Shaped pulses, adiabatic rapid passage, and composite sequences use tailored time-dependent interactions to faithfully implement the desired quantum operations while compensating for system inhomogeneities\cite{fang2018improving,kovachy2012adiabatic,cummins2000use}.

For ultracold atomic ensembles with narrow velocity distributions, rectangular pulses can achieve high-fidelity population transfer. In contrast, for thermal clouds with broader velocity spreads, fixed-amplitude rectangular pulses cannot effectively compensate for Doppler-induced frequency shifts, thereby limiting their applicability\cite{lopez2024impact,fang2018improving,wang2024amplitude}.

To clearly isolate and study the core impact of pulse shaping on interferometer robustness, this work employs a simplified two-level atomic model. We acknowledge that the Raman transition in a real atomic system, such as $^{87}$Rb, is a three-level process. The necessary adiabatic elimination of the intermediate state introduces additional complexities and velocity-selective effects not captured by our model. However, the two-level approximation is a standard and effective approach for foundational studies, allowing for a clear, comparative analysis of how different pulse shapes perform against the primary error source of detuning caused by thermal velocity spread.

In this work, we systematically investigate the use of super-Gaussian pulses as beam-splitting and mirror pulses to enhance robustness. By comparing rectangular, Gaussian, and super-Gaussian pulses of orders 2 through 10 under conditions of pulse area and detuning errors, we focus on their relative performance, particularly in thermal atomic clouds, we also observe that the fringe contrast enhancement provided by super-Gaussian pulses saturates beyond a certain order, indicating a practical optimization limit. Our analysis reveals that while the peak performance of high-order super-Gaussian pulses is comparable to Gaussian pulses at very low temperatures, their key advantage lies in their enhanced robustness across a wider range of thermal conditions. Notably, under such conditions, replacing a rectangular pulse sequence with a 4th-order super-Gaussian sequence can improve the fringe contrast from 0.0895 to 0.1709. These results highlight that super-Gaussian pulses offer an effective route toward mitigating contrast degradation caused by thermal motion in atom interferometers.

\section{Theoretical framework}

Consider a two-level atomic system, where the internal states $\left|0\right\rangle$ and $\left|1\right\rangle$ are subject to an external driving field. In the rotating frame, the Hamiltonian of the system can be written as\cite{saywell2018optimal}
\begin{equation}
H = \frac{\hbar}{2} \left( \delta \sigma_z + \Omega e^{i\phi_t} \sigma_+ + \Omega e^{-i\phi_t} \sigma_- \right) \label{equation1}
\end{equation}
here, $\sigma_z$ is the Pauli spin operator, and $\sigma_+$ and $\sigma_-$ are the raising and lowering operators, respectively, which describe the evolution of the two-level system in its matrix representation. The term $\Omega$ denotes the Rabi frequency, $\phi_t$
is the phase of the driving field, and $\delta$ is the detuning, defined as\cite{berman1997atom}
\begin{equation}
\delta(t) = \omega_L - \omega_{12} - \left( \frac{\mathbf{p} \cdot \mathbf{k}}{m} + \frac{\hbar |\mathbf{k}|^2}{2m} + \mathbf{k} \cdot \mathbf{g}t \right) + \delta^{\text{AC}} \label{equation2}
\end{equation}
Here, $\omega_L$ is the frequency of the driving field, $\omega_{12}$ is the transition frequency of the two-level system, and $\mathbf{k}$ is the wave vector. The term $\delta^{\text{AC}}$ accounts for the AC Stark shift. The Doppler shift term $(\mathbf{p} \cdot \mathbf{k}) \, / \, m$ reflects the effect of the motion of the atoms, while $\hbar |\mathbf{k}|^2 / 2m$ describes the recoil-induced correction due to photon momentum exchange. In interferometry, the additional frequency shift $\mathbf{k} \cdot \mathbf{g}t$ due to gravity is an important factor affecting the phase evolution of the system.

\begin{figure}[H]
\begin{adjustwidth}{2cm}{-0cm}
\includegraphics[scale=0.5]{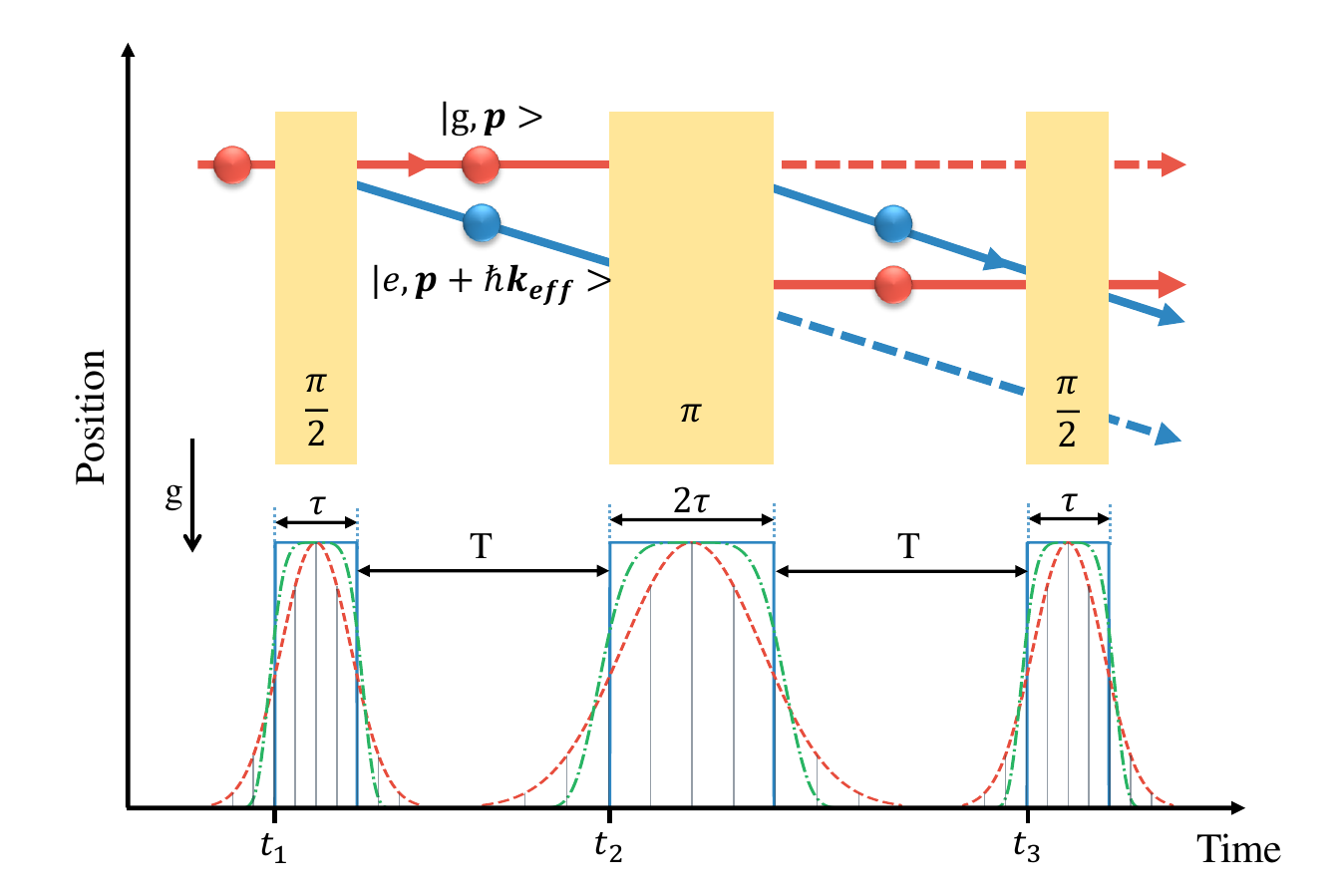}
\end{adjustwidth}
\caption{Principle of Atom Interferometry in MZ configuration. Top: The initial state of the atomic cloud is $|1\rangle$. The $\frac{\pi}{2}$ pulse acts as a splitter, causing the atom to split into a superposition of two atomic states $|0\rangle$ and $|1\rangle$. Due to the increased momentum of the $|1\rangle$ state atom, it follows a different path. Subsequently, after a time $T$, a $\pi$ pulse is applied, causing the two atomic states to exchange. Finally, after another time $T$, the second $\frac{\pi}{2}$ pulse recombines the atomic states to produce an interference signal. Bottom: Conventional rectangular, Gaussian, and super-Gaussian Pulses sequences. In numerical modeling, a piecewise constant approximation method is used to represent them as $N$ rectangular units of equal width.\label{figure1}}
\end{figure}  

Figure \ref{figure1} shows the interaction between the laser pulse and the atoms in the Mach Zender(MZ) atom interferometer. The paths shown by their dashed lines at the top are parasitic paths due to imperfect interferometers, which are usually larger than the Raman line widths and also cause a reduction in the contrast of the interference fringes.

The time evolution of the system is governed by the propagator $U(t)$, which is expressed as\cite{stoner2011analytical}
\begin{equation}
U(t) = \exp \left[ -\frac{i}{\hbar} \int_0^t H(t') dt' \right] \label{equation3}
\end{equation}

For the action of a pulse over a continuous time interval $\Delta t$, the propagator can be written as\cite{dunning2014composite}
\begin{equation}
U(\Delta t) = \begin{pmatrix}
C & -iS^* \\
-iS & C \label{equation4}
\end{pmatrix}
\end{equation}
where
\begin{equation}
C = \cos(\theta / 2) + i (\delta / \Omega_R) \sin(\theta / 2) \label{equation5}
\end{equation}
\begin{equation}
S = e^{i \phi} (\Omega_{\text{eff}} / \Omega_R) \sin(\theta / 2) \label{equation6}
\end{equation}
Here, $\Omega_R = \sqrt{\delta^2 + \Omega_{\text{eff}}^2}$. $\Omega_{\text{eff}} = \Omega_1^* \Omega_2 / 2\Delta$ represents the effective Rabi frequency between the energy levels $|g, \mathbf{p}\rangle$ and $|e, \mathbf{p} + \hbar \mathbf{k}_{\text{eff}}\rangle$\cite{luo2016contrast}, where $\Omega_1$ is the laser $w_1$ coupling to the energy levels $|g, \mathbf{p}\rangle$ and $|i, \mathbf{p} + \hbar \mathbf{k}_1\rangle$, and $\Omega_2$ is the laser $w_2$ coupling to $|i, \mathbf{p} + \hbar \mathbf{k}_1\rangle$ and $|e, \mathbf{p} + \hbar \mathbf{k}_{\text{eff}}\rangle$. The angle $\theta$ represents the rotation angle of the state vector around the Bloch sphere, calculated as $\theta = \Omega_R \Delta t$.

\begin{figure}[H]
\begin{adjustwidth}{-2cm}{-2cm}
\centering
\subfloat[\centering]{\includegraphics[scale=0.35]{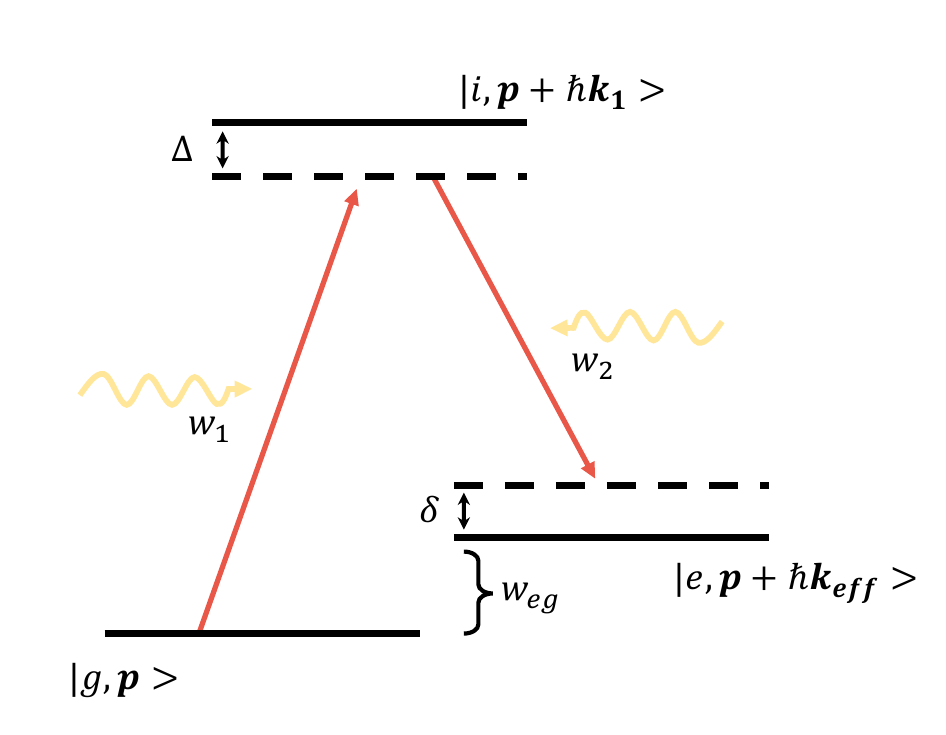}}
\subfloat[\centering]{\includegraphics[scale=0.3]{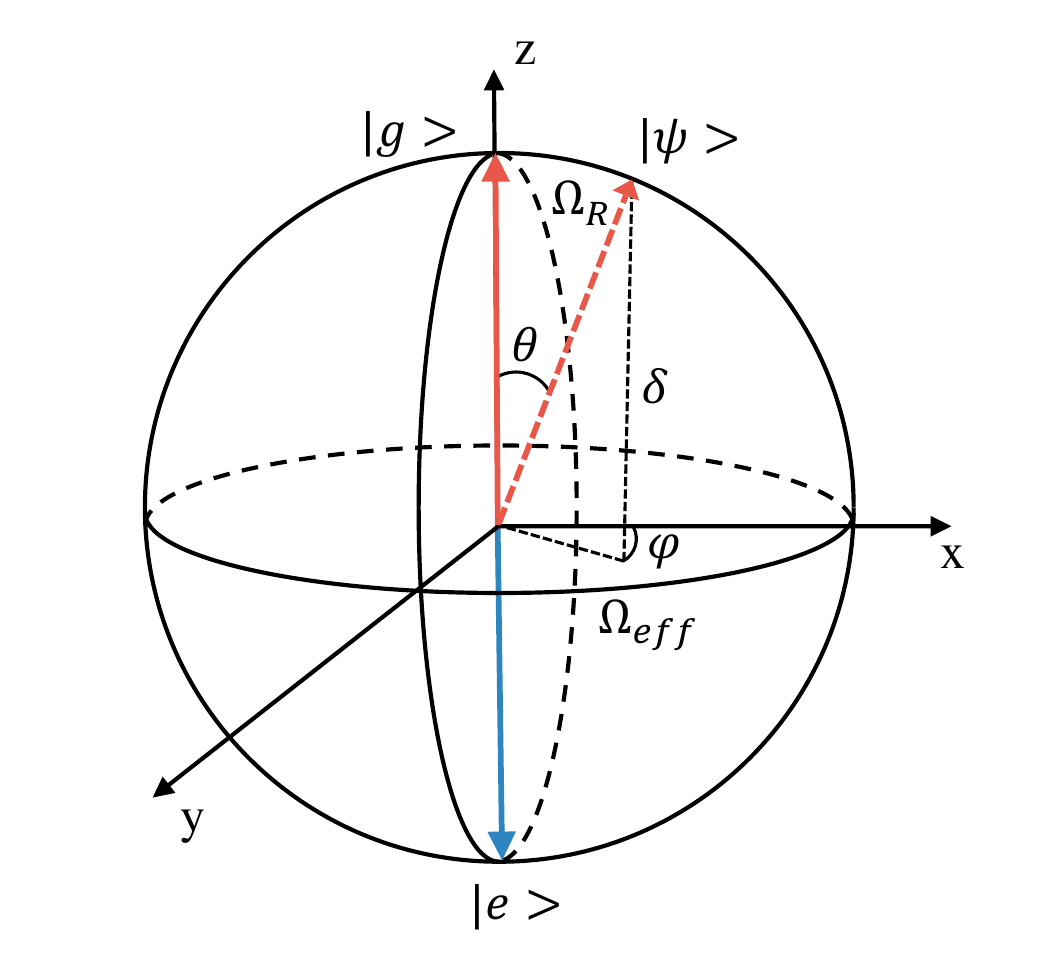}}\\
\end{adjustwidth}
\caption{(\textbf{a}) The system of interaction between the optical field and atoms during the double-photon Raman process. (\textbf{b}) Bloch sphere representation of the atomic interference process.\label{figure2}}
\end{figure}

The visualization of the atomic state evolution process can also be represented on the Bloch sphere\cite{torosov2019arbitrarily}. The atomic cloud quantum state during the interference process can be described on the Bloch sphere as follows\cite{shore2011manipulating}:
\begin{equation}
|\psi\rangle = \cos\left(\frac{\nu}{2}\right)|0\rangle + e^{i\varphi}\sin\left(\frac{\nu}{2}\right)|1\rangle \label{equation7}
\end{equation}
Here, $\nu$ and $\varphi$ represent the polar and azimuthal angles on the Bloch sphere, respectively, whose variations determine the dynamic evolution of the two-level atomic system. Any superposition state between $|g, \mathbf{p}\rangle$ and $|e, \mathbf{p} + \hbar \mathbf{k}_{\text{eff}}\rangle$ can be described by a point on the Bloch sphere. Figure \ref{figure2}(b) illustrates the interference path represented on the Bloch sphere, where the angular velocity on the Bloch sphere is given by the parameters of the driving field, satisfying the equation
\begin{equation}
\frac{d\mathbf{S}}{dt} = \boldsymbol{\Omega}(t) \times \mathbf{S} \label{equation8}
\end{equation}
where $\mathbf{S} = (\langle \sigma_x \rangle, \langle \sigma_y \rangle, \langle \sigma_z \rangle)$ represents the Bloch vector, and $\boldsymbol{\Omega}(t) = (\Omega \cos \phi, \Omega \sin \phi, \delta)$ describes the effective driving force of the system. The Bloch vector rotates around the rotation axis on the Bloch sphere at an angular velocity $\Omega_R$. If the duration of the driving laser field is $\tau$, the Bloch vector will rotate by an angle $\theta = \Omega_R \tau$.

Based on the above description, the propagator can be expressed in the form $U(\theta, \phi, \alpha)$, which represents a general unitary rotation on the Bloch sphere, where $\phi$ represents the instantaneous phase of the Raman laser field, and $\alpha$ denotes the field vector's precession angle. This signifies the relationship between the initial and final states of the atom $|1\rangle = U |0\rangle$. Considering the standard MZ interference pulse sequence acting on the atom interferometry, the final atomic state at the output can be clearly described by the corresponding evolution operator and the free evolution operator between the pulses for the three typical pulse sequences:

\begin{equation}
\begin{bmatrix}
c_e(t_3 + \tau_3) \\
c_g(t_3 + \tau_3)
\end{bmatrix}
=
U_{\pi/2} U_{F1} U_{\pi} U_{F2} U_{\pi/2}
\begin{bmatrix}
c_e(0) \\
c_g(0) \label{equation9}
\end{bmatrix}
\end{equation}
Here, $U_{\pi/2}$ and $U_{\pi}$ represent the pulse propagation operators corresponding to $\pi/2$ and $\pi$ pulses, respectively, while $U_{\text{Fj}} (j=1,2)$ describes the free evolution of atoms in the absence of external fields between pulses (the free evolution operator $U_{\text{F}}$ can be calculated by setting $\Omega_{\text{eff}} = 0$ in the propagator). To quantify the performance of these quantum operations, we use the term fidelity, defined as the squared overlap between the final state($|\psi_{final}\rangle$) and the ideal target state($|\psi_{target}\rangle$), i.e., $F = |\langle \psi_{target} | \psi_{final} \rangle|^2$. An ideal, perfect $\pi/2$ pulse aims to achieve a transition probability of 0.5, creating an equal superposition of the two atomic states. This corresponds to a fidelity of 1.0 with the target superposition state. The $\pi$ pulse aims for a complete population inversion (a transition probability of 1.0), also corresponding to a fidelity of 1.0 with the excited target state. During the free evolution period $T_j$, the excited state acquires a phase $\exp(i\Phi_j/2)$, and the ground state acquires a phase $\exp(-i\Phi_j/2)$, where $\Phi_j = \int_0^T dt' \delta(t')$ represents the total accumulated phase due to the interaction-free period\cite{stoner2011analytical}.

During the interference, the freely falling atomic cloud experiences a velocity shift due to gravitational acceleration, resulting in a Doppler shift. To maintain resonance between the laser and the center of the atomic cloud, a linear chirp technique is typically employed, where the Raman laser frequency is modulated linearly in time to compensate for this global motion-induced frequency shift\cite{Peters_1999}. However, due to the thermal motion within the atoms, there is an inherent velocity distribution of the atomic cloud, and there is some deviation in the velocities of individual atoms with respect to the center of the atomic cloud $v$. This velocity spread introduces a local detuning $\delta_v = |\mathbf{k}_{\text{eff}} \cdot \mathbf{v}|$ (equivalent to $(\mathbf{p} \cdot \mathbf{k}) \, / \, m$ in equation (\ref{equation2})), which reduces the driving efficiency of laser pulses for atoms with different velocities, consequently degrading the interference fringe contrast. Although the linear chirp can effectively compensate the Doppler shift due to the falling motion, the effective detuning felt by atoms with different velocities is different due to the Maxwell-Boltzmann velocity distribution of the atomic cloud. This detuning distribution leads to an increase in phase randomness during the evolution of quantum states, producing a phase diffusion effect, which is thus one of the main factors limiting the contrast enhancement of interferometers.

By establishing an ordered interaction between pulse propagators and applying them to atoms initially in the ground state, an analytic expression for the excited state population can be derived. Specifically for the $\pi/2$-$\pi$-$\pi/2$ sequence, the transition probability has been mathematically formulated by Stoner et al.\cite{stoner2011analytical,butts2011light}:
\begin{equation}
\begin{split}
P_e &= |S_1|^2 |S_2|^2 |S_3|^2 + |C_1|^2 |S_2|^2 |C_3|^2 + |S_1|^2 |C_2|^2 |C_3|^2 + |C_1|^2 |C_2|^2 |S_3|^2 \\
&\quad  - 2 \text{Re} \left[ \exp(i \phi_p) C_1 S_1 (S_2^*) ^2 C_3^* S_3 \right] \label{equation10}
\end{split}
\end{equation}
where \( C_i \) and \( S_i \) are defined in equation (\ref{equation5}) and (\ref{equation6}). To further simplify the derivation, we consider identical initial and final pulses by setting \( S_1 = S_3 \) and \( C_1 = C_3 \). By applying this condition to equation (\ref{equation10}) and then collecting the terms that are constant with respect to the interferometric phase $\Phi_g$ into a background term $P_0(\delta)$ and the oscillating terms into a cosine function with amplitude B, the transition probability at the interferometer output can be reduced to the following canonical form:

\begin{equation}
P_e = \frac{1}{2} \left\{ P_0(\delta) + B \cos \left[ \Phi_g + \phi(\delta) \right] \right\} \label{equation11}
\end{equation}
where \(\Phi_g\) represents the accumulated interferometric phase of atoms throughout the entire pulse sequence, \(\phi(\delta)\) characterizes the relative phase shift introduced by the pulses, given by \(\phi(\delta) = \phi_{\pi/2}^{S1} + \phi_{\pi/2}^{S3} - 2\phi_{\pi}^{S2}\), the parameters $P_0(\delta)$ is an offset and B is the amplitude of the oscillatory function. Under ideal conditions, this phase term can be simplified to \(\alpha T^2\), where $\alpha$ is the linear chirp rate of the Raman laser frequency, applied to compensate for the gravitational acceleration, and T is the free evolution time. After performing the thermal averaging over the velocity distribution, the final transition probability can be expressed as:
\begin{equation}
P_e = \int_{-\infty}^{\infty} \frac{1}{2} \left\{ P_0(\delta) + B \cos \left[ \Phi_g + \phi(\delta) \right] \right\} f(\delta) \, d\delta \label{equation12}
\end{equation}

Therefore, the longitudinal velocity distribution of the atomic cloud, horizontal expansion effects, and the intensity inhomogeneity of Raman beams in the horizontal direction will all degrade the performance of propagation operators, leading to attenuation of interference fringe contrast. Experimentally, by adjusting the Rabi frequency and phase of each pulse, the interference performance can be significantly optimized.

Specifically, \(\Phi_g\) is determined by the phase accumulation from each Raman pulse, depending on both the duration of pulse interaction and the coupling strength between atoms and the laser field. The contrast and fringe morphology are influenced by detuning and bias effects. Consequently, optimized pulse adjustment -- particularly in terms of pulse shaping and phase optimization -- constitutes a crucial factor for enhancing overall interference performance, with experimental results from numerous labs having validated its effectiveness\cite{saywell2023enhancing,zhao2022optimized,dedes2023optimizing}.

\section{Pulse Performance}
In the MZ sequence of cold-atom interferometers, the dynamical response of $\pi/2$ and $\pi$ pulses depends critically on the temporal evolution characteristics of Raman laser pulses. This study systematically evaluates the robustness differences among rectangular, Gaussian, and super-Gaussian pulse sequences under composite noise environments involving large detuning and pulse length error. In our numerical simulations, the pulse profiles for Gaussian and $n^{th}$-order super-Gaussian pulses are expressed as\cite{karar2019super}:
\begin{equation}
f(t) = \exp\left(-\left(\frac{t^2}{2\zeta^2}\right)^n\right) \label{equation13}
\end{equation}
where $\zeta$ denotes the standard deviation, and the order $n$ of the super-Gaussian pulse is set within the range of 2 to 10. When the order $n=1$, equation (\ref{equation13}) reduces to the standard Gaussian pulse profile.

\begin{figure}[!htbp]
\begin{adjustwidth}{0.5cm}{1cm}
\centering
\subfloat[\centering]{\includegraphics[scale=0.55]{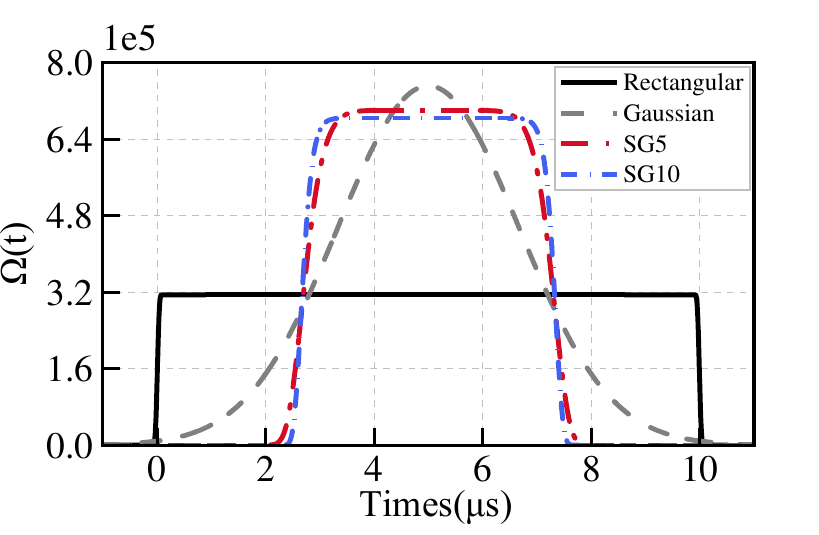}}
\subfloat[\centering]{\includegraphics[scale=0.55]{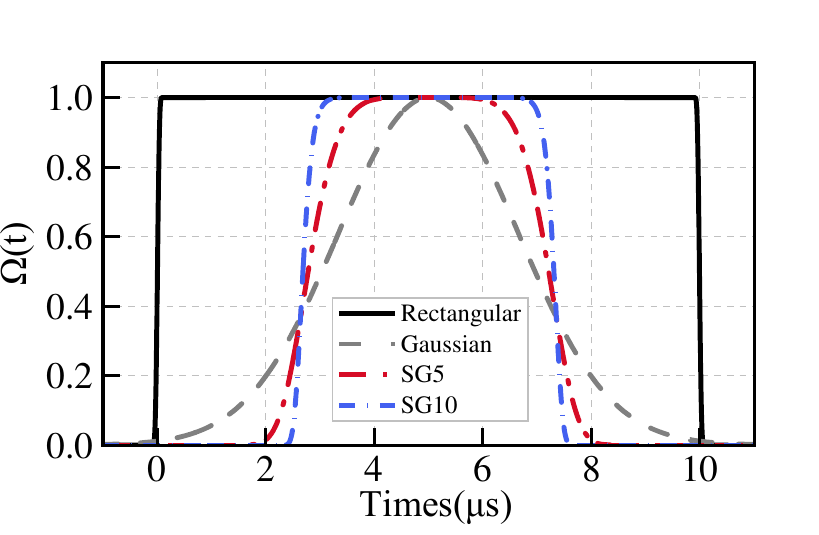}}
\end{adjustwidth}
\caption{(\textbf{a}) Temporal evolution comparison of effective Rabi frequencies for rectangular, Gaussian, and super-Gaussian ($n=5, 10$) pulses. (\textbf{b}) Normalized pulse intensity profiles, with each pulse normalized to its peak amplitude. \label{figure3}}
\end{figure} 

The super-Gaussian pulse, as a typical shaped pulse, exhibits a time-domain intensity profile intermediate between standard Gaussian pulses and pure flat-top pulses. Its defining characteristic is a flat-top with steep, yet smoothly-varying edges, which are crucial for minimizing off-resonant excitations and suppressing the high-frequency noise that is inherent in the abrupt transitions of idealized rectangular pulses. This engineered temporal profile is key to its enhanced robustness. 

Experimentally, the precise control over the pulse shape is achieved using a high-resolution arbitrary waveform generator (AWG). The AWG is used to modulate the RF driving power of a commercially available acousto-optic modulator (AOM). The diffraction efficiency of the AOM is directly proportional to the RF power, allowing for a precise mapping of the time-dependent voltage signal from the AWG to the effective Rabi frequency of the laser beam. This enables us to generate any arbitrary temporal pulse shape, including the super-Gaussian profile. The fidelity of this process depends on the sampling rate and bit depth of the AWG and the bandwidth of the AOM\cite{Zhao_2020}. The temporal discretization method used in our numerical simulations  is a direct analogue to the piecewise-constant approximation used by an AWG to generate a continuous waveform in an actual experiment, where a high sampling rate provides an excellent approximation. The amplitude of super-Gaussian pulses follows\cite{beirle2017parameterizing}:
\begin{equation}
A = A_0 \exp\left(-\left(\frac{t^2}{2\zeta^2}\right)^n\right)  \label{equation14}
\end{equation}
where $A_0$ denotes the peak pulse amplitude and $n$ represents the order of the super-Gaussian function. The super-Gaussian pulse model achieves continuous tunability of pulse shapes through the order parameter $n$. When $n=1$, the function reduces to describe standard Gaussian pulse amplitudes, maintaining compatibility with conventional Gaussian pulse frameworks while providing an additional degree of freedom —the super-Gaussian order, n—which allows for the continuous tuning of the pulse shape from a standard Gaussian (for n=1) towards a flat-top profile. This tunability is the key to optimizing robustness. Compared to Gaussian pulses, the super-Gaussian extension incorporates the shape parameter $n$ to characterize diverse profiles, constituting a remarkably simple yet powerful generalization.

For super-Gaussian and Gaussian pulses with time-varying amplitudes, we employ a temporal discretization method to study their interaction with atoms. The pulse duration $\tau$ is divided into $N$ equal time steps $\Delta t = \tau/N$, forming a sequence of rectangular sub-pulses as shown in the lower part of figure \ref{figure1}. Each segment maintains a constant amplitude, and when the discretization $N \to \infty$, the overall effect becomes equivalent to that of an amplitude-modulated pulse. This discretization strategy effectively implements equivalent amplitude-modulation modeling for the interaction between time-varying pulses and atomic systems. Even for complex pulse shapes, sufficiently dense piecewise-constant sampling can achieve excellent approximation. Within this time-segmented approximation framework, the total system propagator can be decomposed into an ordered product of $N$ discrete rectangular sub-pulse propagators:

\begin{equation}
U = \prod_{k=1}^{N} U_k(\Delta t) = U_N U_{N-1} U_{N-2} \cdots U_3 U_2 U_1 \label{equation15}
\end{equation}
where $U_k$ represents the unitary propagator corresponding to the $k^\text{th}$ pulse segment. When replacing rectangular pulses with amplitude-modulated pulses, the pulse area conservation condition must be strictly satisfied:

\begin{equation}
A = \int_{-\infty}^{\infty} \Omega_{\text{eff}}(t) \, dt = \text{const} \label{equation16}
\end{equation}

For $\pi/2$ pulses, the conserved area must satisfy $A = \pi/2$, while $\pi$ pulses require $A = \pi$. This area conservation condition ensures coherent superposition of atomic wavepackets and complete population transfer, respectively.

In numerical simulations, the maximum Rabi frequency of rectangular pulses is set as $\Omega_{\text{rec}} = \pi \times 10^5\,\text{kHz}$. The effective Rabi frequencies of rectangular and Gaussian pulses used for calculations are shown in figure \ref{figure3}(a). Here, the pulse area is strictly conserved to ensure coherent population transfer, but the peak amplitude varies for different pulse shapes. figure \ref{figure3}(b) shows the pulse shapes normalized to their peak amplitude for a direct visual comparison of their temporal profiles. Specifically, the displayed pulses are $\pi$-pulses with 10$\mu$s duration. In an idealized cold-atom interferometer model where Raman pulses exhibit uniform transverse intensity distribution and the atomic cloud maintains ultracold temperature with negligible initial velocity spread, all atoms would experience identical effective Raman pulse areas in this limit. However, in real cold-atom systems, the atomic cloud undergoes free expansion during free fall due to residual transverse velocities. According to statistical mechanics principles, the expanding cloud develops Gaussian phase-space distribution characteristics, where position $\mathbf{r} = (x,y,z)$ and velocity $\mathbf{v} = (v_x,v_y,v_z)$ remain statistically independent. The phase-space probability distribution can be expressed as\cite{brzozowski2002time}:

\begin{equation}
N(x,y,z,v_{x},v_{y},v_{z}) = \prod_{\mu \in \{x,y,z\}} f(\mu_{0},\zeta _{0}) f(v_{\mu_{0}},\zeta_{v}) \label{equation17}
\end{equation}
where $f$ denotes the one-dimensional Gaussian distribution. The relative velocity component $v$ of the atomic cloud along the wavevector direction also follows a 1D Gaussian distribution:

\begin{equation}
f(v_z, \mu_0, \zeta_z) = \frac{1}{\sqrt{2\pi\zeta^2}} \exp\left[ -\frac{(v_z - \mu_0)^2}{2\zeta^2} \right], \label{equation18}
\end{equation}
Here $\mu_0$ represents the central value of the velocity distribution. This velocity distribution is the source of the primary decoherence mechanism studied in this work, as each velocity component $v_z$ for an individual atom creates a Doppler-induced detuning $\delta_v = k_{\text{eff}} v_z$, leading to a distribution of detunings across the atomic cloud.
This distribution couples with the effective wavevector $k_{\text{eff}}$ to produce the two-photon detuning $k_{\text{eff}}f(v_z, \mu_0, \zeta_z)$, where the resultant detuning is directly correlated with the atomic temperature. In our simulations, the 'temperature' parameter refers to the temperature of the entire, unselected atomic cloud. Therefore, it characterizes the width of the full thermal velocity distribution in the longitudinal direction, rather than a selected subset of atoms.

\subsection{Measures of pulse performance}
\begin{figure}[!htbp]
\begin{adjustwidth}{1.7cm}{1cm}
\includegraphics[scale=0.45]{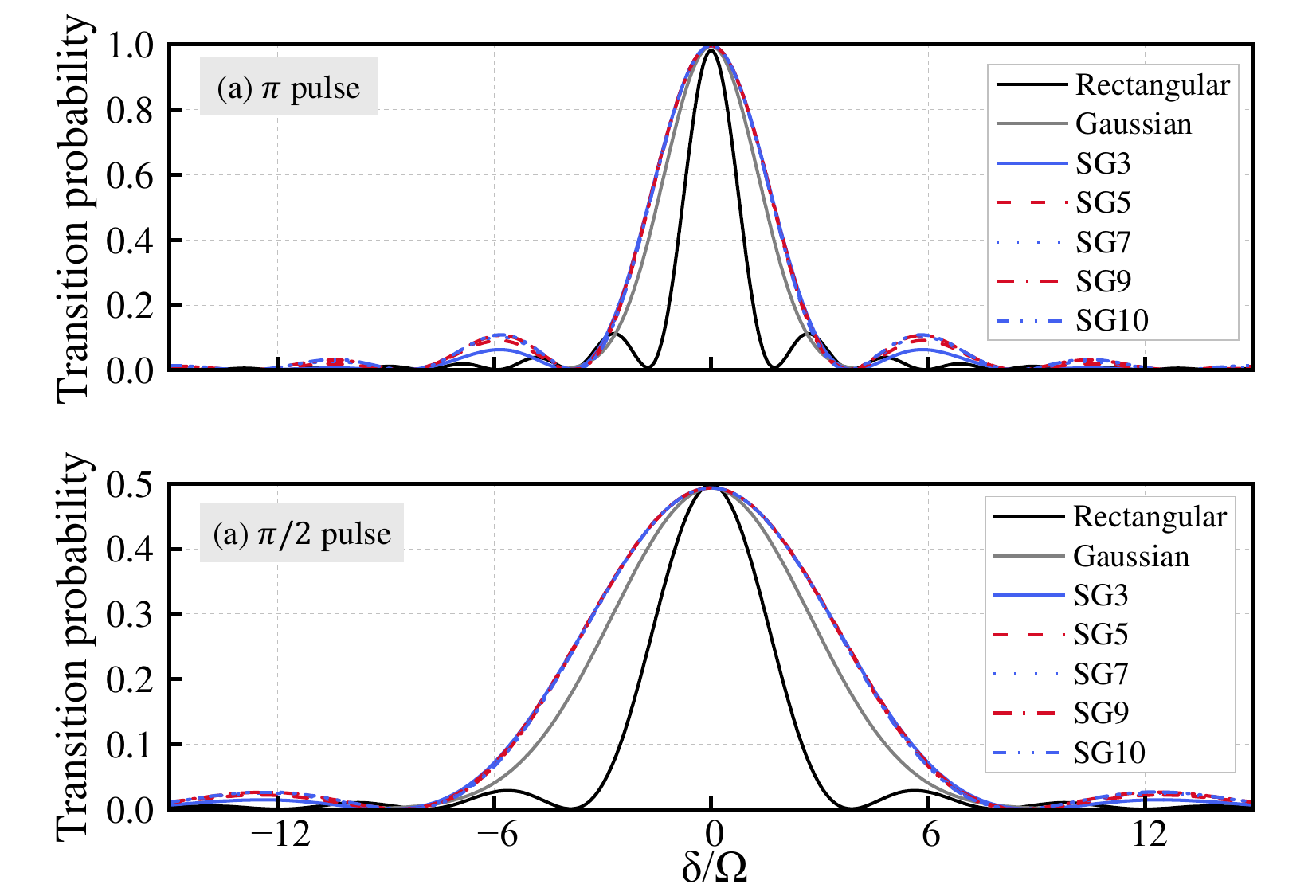}
\end{adjustwidth}
\caption{The detuning-dependent transfer probability as a function of temporal pulse profiles. The response curves are shown for rectangular pulses, Gaussian pulses, and odd-order super-Gaussian pulses ($n=3,5,7,9,10$). Top: Transition probability under $\pi$ pulses. Bottom: Transition probability under $\pi/2$ pulses.\label{figure4}}
\end{figure}

Using rectangular $\pi$ and $\pi/2$ pulses (black curves in figure \ref{figure4}) as a reference benchmark, we note that standard rectangular pulse excitations exhibit poor tolerance to Doppler broadening in atomic clouds, i.e., low robustness against detuning. Under significant two-photon detuning conditions, rectangular pulses fail to effectively perform beam splitting or quantum state swapping. In contrast, Gaussian-family pulses demonstrate superior detuning tolerance, enabling optimized atomic population inversion, beam-splitting efficiency, and quantum state transfer fidelity under substantial detuning. 

In this work, we use the term detuning tolerance to quantify the robustness of each pulse shape against detuning errors. Detuning tolerance is defined as the range of detuning over which the pulse maintains a transition fidelity above a specified threshold, such as 0.5 or 0.9. This parameter is directly derived from the transition probability curves shown in Figure 4, where the range is measured at the designated fidelity levels.

Super-Gaussian pulses outperform Gaussian pulses in both $\pi$ and $\pi/2$ pulse regimes, providing enhanced response to broader atomic detuning ranges. Although high-order super-Gaussian pulses visually resemble ideal rectangular pulses, their key physical advantage lies in their smoothly-varying temporal edges(see figure \ref{figure3}(b)). Unlike the idealized, infinitely steep edges of a perfect rectangular pulse (which is physically impossible to generate and can introduce high-frequency noise), super-Gaussian pulses maintain a smooth, albeit steep, transition. The abrupt, discontinuous edges of an ideal rectangular pulse in the time domain necessarily create strong, high-frequency components in its spectrum, which manifest as the large side lobes. These side lobes lead to unwanted off-resonant excitations, degrading the pulse fidelity for non-zero detuning. In contrast, the temporal smoothness of the Gaussian and super-Gaussian pulses confines their energy to a narrower frequency band, effectively suppressing the side lobes. This is the physical reason for their enhanced robustness against detuning: they interact strongly only with atoms near resonance. This smooth profile is crucial for suppressing phase noise and off-resonant excitations, which are particularly detrimental in cold atomic clouds. Therefore, while a high-order super-Gaussian pulse may appear similar to a rectangular pulse, its engineered smooth edges provide a tangible and significant benefit in noise resistance and robustness, as demonstrated by our simulation results.\cite{fang2018improving,Gillen_Christandl_2016,cheinet2008measurement}.

Investigating whether there exists an optimization limit for super-Gaussian pulse waveforms is a critical issue in cold-atom interferometry. Table \ref{table1} presents the fidelity of atomic interactions driven by $\pi$ and $\pi/2$ pulses with different shapes. The Gaussian-family pulses exhibit stronger detuning robustness compared to rectangular pulses. For super-Gaussian pulses, varying the order $n$ from 2 to 10 induces only minor fidelity variations. Specifically, the fourth-order super-Gaussian $\pi$ pulse achieves the maximum detuning tolerance of 1.74520 at fidelity $>0.5$ and 0.68023 at fidelity $>0.9$. Similarly for $\pi/2$ pulses, the fourth-order super-Gaussian configuration maintains superior performance with maximum detuning tolerances of 3.79894 ($>0.5$ fidelity) and 1.43993 ($>0.9$ fidelity). These results demonstrate that the variation in pulse order affects the fidelity of super-Gaussian pulses only at the order of magnitude of $1\text{e}^{-2}$.

\begin{table}[!htbp]
\centering
\caption{Comparison of fidelity among rectangular, Gaussian, and super-Gaussian pulses. The atomic cloud temperature is set to 3 $\mu$K without vertical velocity selection, and the initial radius is 1.5 mm. The $\pi$-pulse duration $\tau_{\pi}$ is 10 $\mu$s, while the $\pi/2$-pulse duration $\tau_{\pi/2}$ is 5 $\mu$s. The laser beam radius is 10 mm.\label{table1}}
\begin{tabularx}{\linewidth}{>{\centering\arraybackslash}X >{\centering\arraybackslash}X >{\centering\arraybackslash}X >{\centering\arraybackslash}X >{\centering\arraybackslash}X >{\centering\arraybackslash}X}
\toprule
\multirow{2}{*}{$\pi$ Pulse} & \multicolumn{2}{c}{Detuning tolerance} & \multirow{2}{*}{$\pi$ / 2 Pulse} & \multicolumn{2}{c}{Detuning tolerance} \\
\cmidrule(lr){2-3} \cmidrule(lr){5-6}
& Fidelity $>$0.5 & Fidelity $>$0.9 & & Fidelity $>$0.5 & Fidelity $>$0.9 \\
\midrule
Rec & 0.80691 & 0.31231 & Rec & 1.73225 & 0.68879 \\
Gaussion & 1.49807 & 0.58407 & Gaussion & 3.16418 & 1.17089 \\
SG2 & 1.74140 & 0.67967 & SG2 & 3.70234 & 1.39388 \\
SG3 & 1.74488 & 0.67984 & SG3 & 3.78724 & 1.43263 \\
SG4 & 1.74520 & 0.68023 & SG4 & 3.79894 & 1.43993 \\
SG5 & 1.74488 & 0.68005 & SG5 & 3.79454 & 1.43963 \\
SG6 & 1.74441 & 0.68050 & SG6 & 3.78634 & 1.43723 \\
SG7 & 1.74396 & 0.68003 & SG7 & 3.77794 & 1.43423 \\
SG8 & 1.74352 & 0.68054 & SG8 & 3.76974 & 1.43173 \\
SG9 & 1.73315 & 0.67998 & SG9 & 3.76264 & 1.42913 \\
SG10 & 1.73282 & 0.67995 & SG10 & 3.75654 & 1.42683 \\
\bottomrule
\end{tabularx}
\end{table}

\begin{figure}[!htbp]
\begin{adjustwidth}{2.5cm}{2cm}
\includegraphics[scale=0.45]{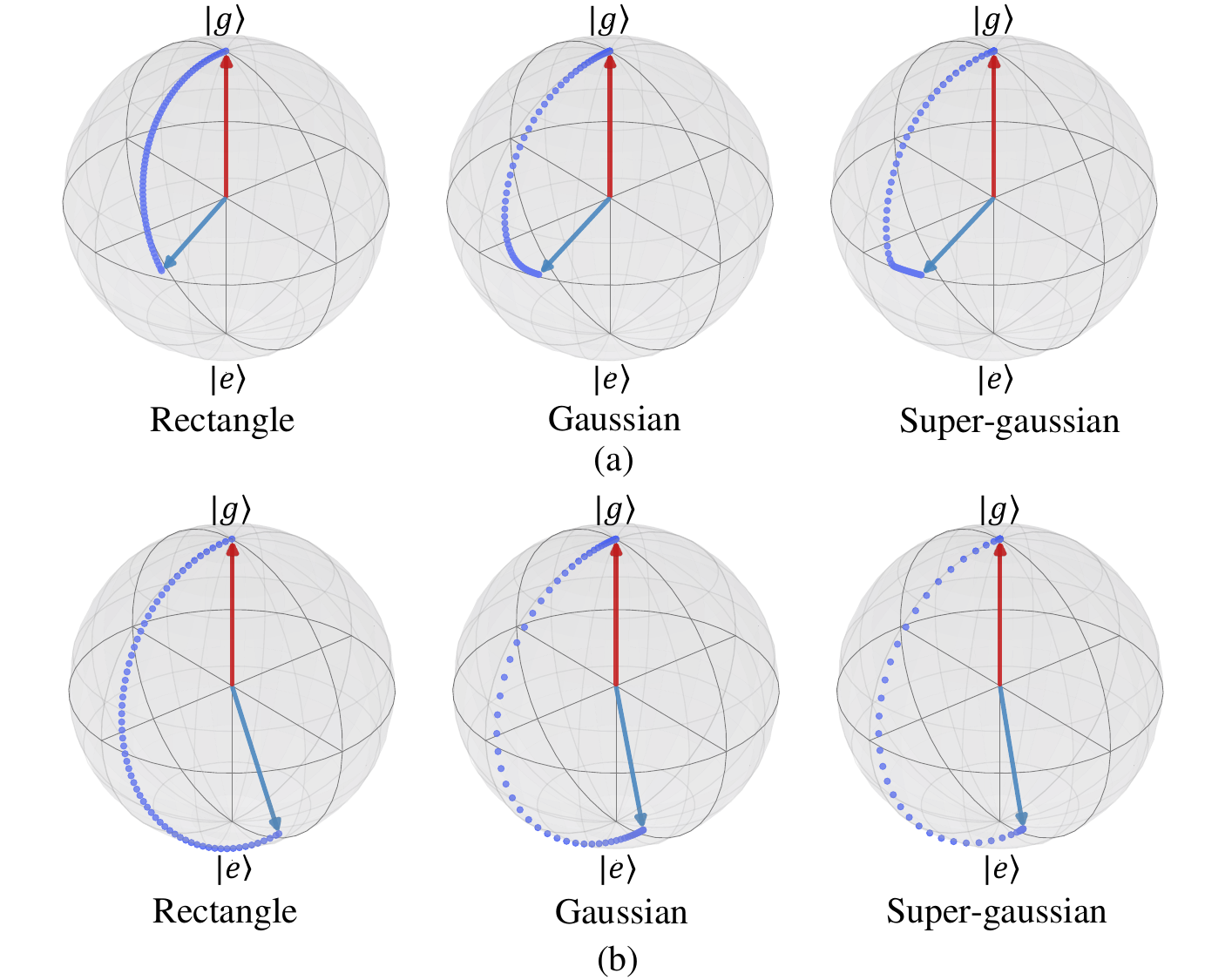}
\end{adjustwidth}
\caption{(a) Quantum state evolution trajectories on the Bloch sphere under rectangular, Gaussian, and super-Gaussian ($n=5$) $\pi/2$ pulses with two-photon detuning $\delta = +0.5\Omega_{\text{rec}}$. (b) Evolution trajectories under rectangular, Gaussian, and super-Gaussian ($n=5$) $\pi$ pulses with $\delta = +0.2\Omega_{\text{rec}}$. The initial state is $|g\rangle$ for all cases, where the ideal $\pi/2$ pulse targets the state $\frac{1}{\sqrt{2}} \left( |g\rangle - i |e\rangle \right)$, and the ideal $\pi$ pulse targets the excited state $|e\rangle$.\label{figure5}}
\end{figure}

Visualization of the relationship between fidelity and the shape of interference pulses using the Bloch sphere representation. As shown in figure \ref{figure5}, under rectangular pulses, the quantum state trajectory exhibits significant deviation during rotation about the axis. In contrast, replacing conventional rectangular pulses with super-Gaussian pulses demonstrates smaller errors under large detuning conditions, with the final state approaching closer to the ideal target position. This result indicates superior robustness of super-Gaussian pulses against detuning effects.

Figure \ref{figure5}(a) displays the quantum state evolution trajectories of different pulse shapes for $\pi/2$ pulses in the Bloch sphere representation, with detuning $\delta$ set to $+0.5\Omega_{\text{rec}}$. figure \ref{figure5}(b) shows the evolution trajectories of different $\pi$-pulse shapes under detuning $\delta = +0.2\Omega_{\text{rec}}$ with super-Gaussian pulse order $n=4$. The initial state $\ket{\psi_0}$ and final state $\ket{\psi}$ are denoted by red and blue arrows respectively. For both $\pi/2$ and $\pi$ pulses, the detuning induces rotational deviations of the Bloch vector from ideal positions, generating off-resonance errors.

\begin{figure}[!htbp]
\begin{adjustwidth}{-2cm}{-2cm}
\centering
\subfloat[\centering]{\includegraphics[scale=0.45]{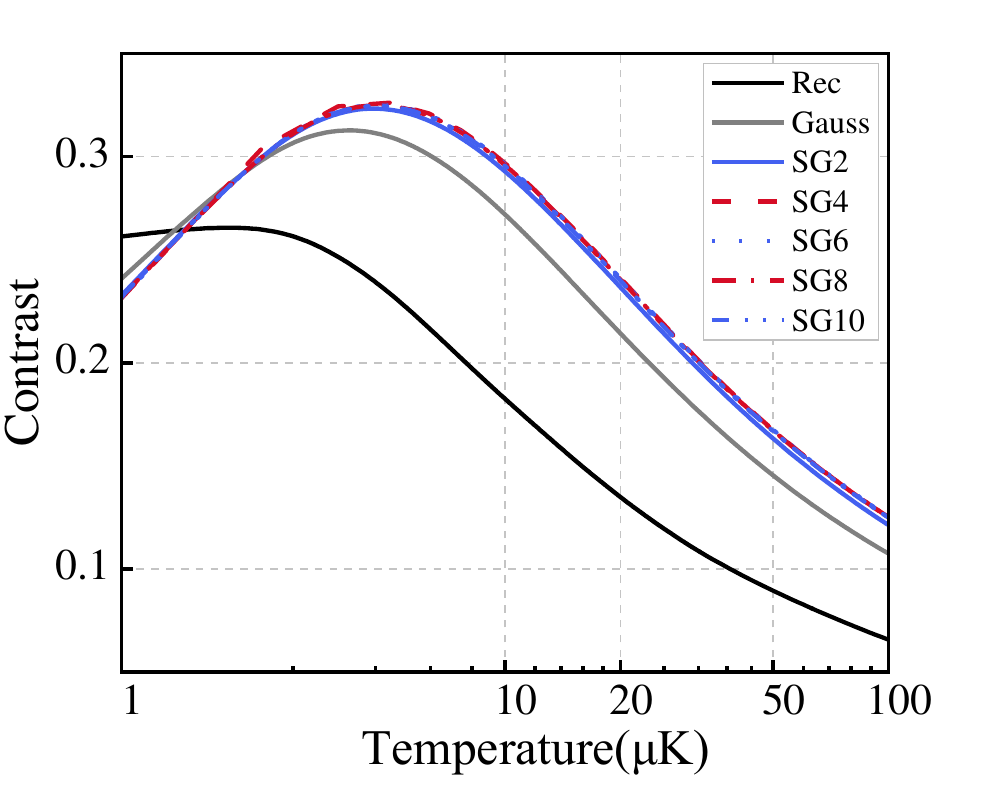}}
\subfloat[\centering]{\includegraphics[scale=0.45]{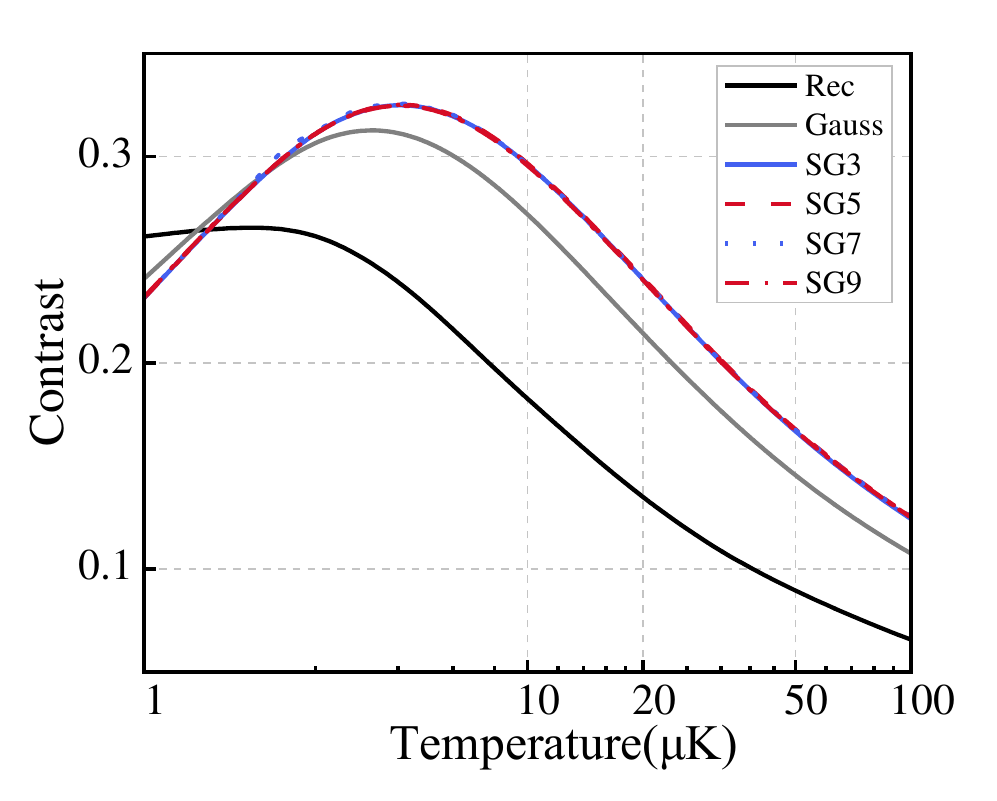}}
\end{adjustwidth}
\caption{Contrast versus temperature for different pulse-shape MZ sequences. All three pulses in the MZ sequence share identical pulse shapes, with the interference fringe contrast plotted as a function of atomic cloud temperature. The laser beam radius in these simulations is 10 mm.\label{figure6}}
\end{figure}

As shown in figure \ref{figure6}, we numerically compared the effects of different MZ pulse sequences on the final contrast of the interferometer for atomic cloud temperatures $T$ ranging from 1~$\mu$K to 100~$\mu$K, while also evaluating the performance of super-Gaussian pulses with orders $n$ varying from 2 to 10. In our simulations, both the $\pi$-pulse and $\pi/2$-pulse durations maintained the same settings as described above, with free evolution times $T_{\text{F1}}$ = $T_{\text{F2}}$ = 100~ms. While Gaussian-family pulses consistently outperformed conventional rectangular pulses, we observed that Gaussian pulse sequences achieved slightly higher interference fringe contrast than super-Gaussian pulses at lower atomic cloud temperatures $(T < 4~\mu\mathrm{K})$. However, the key advantage of the super-Gaussian pulses becomes prominent at higher temperatures, which are more representative of typical experimental conditions without deep cooling or velocity selection. For instance, at $5~\mu\mathrm{K}$, the 4th-order super-Gaussian pulse provides a contrast of 0.1709, a dramatic 90.9\% improvement over the rectangular pulse (0.0895) and a significant 12.2\% improvement over the Gaussian pulse (0.1523), as detailed in Table 2. This demonstrates their superior performance in mitigating contrast degradation caused by a broad thermal velocity distribution.

This phenomenon demonstrates that Gaussian pulses exhibit more significant effects on the coherence of cold atomic clouds at lower temperatures. However, as the temperature increases, the advantages of super-Gaussian pulses become progressively apparent. Particularly at elevated temperatures, super-Gaussian pulses show superior stability and enhanced noise resistance in improving interference fringe contrast. These results indicate that super-Gaussian pulses possess broader applicability for cold atomic systems across different temperature regimes, with their advantages being particularly pronounced in high-temperature environments.

As evident from figure \ref{figure6}, the super-Gaussian pulses show minimal variation in their effect on interference fringe contrast across orders $n$ ranging from 2 to 10, indicating the existence of a saturation point in order optimization. The influence of increasing order on contrast gradually stabilizes, with pulse performance approaching optimal conditions within this order range. Further order enhancement yields diminishing returns, suggesting that the system reaches a performance plateau where additional order increments provide negligible improvement. This behavior likely stems from fundamental physical constraints: higher-order super-Gaussian pulses approach a limiting waveform configuration, potentially due to either the time-bandwidth product limitation or saturation effects in atom-light interaction dynamics\cite{masoudnia2015optimum,chuang1993propagation}.

The low interference contrast observed at very low temperatures (at 1$\mu$K in figure \ref{figure6}) is not primarily due to thermal velocity spread, which is minimal in this regime. Instead, it is dominated by the spatial intensity inhomogeneity of the Raman beams. In our simulations, the finite size of the atomic cloud is exposed to a laser beam with a finite radius. This results in a non-uniform distribution of effective Rabi frequencies and pulse areas across the atomic cloud, causing different atoms to undergo imperfect quantum state manipulations. This effect is a significant source of decoherence that limits the overall fringe contrast, even in the absence of thermal motion.

An interesting feature in figure \ref{figure6} is the non-monotonic behavior of the contrast, which peaks at approximately 4$\mu$K and then slightly decreases at lower temperatures. This suggests the presence of two competing decoherence mechanisms with different temperature dependencies. At higher temperatures (T $>$ 4$\mu$K), the primary limiting factor is the large thermal velocity spread, which causes a wide distribution of Doppler-induced detunings. In this regime, the superior detuning robustness of the shaped pulses, particularly the super-Gaussian, becomes crucial, leading to a significant increase in contrast as the temperature is reduced. Conversely, at very low temperatures (T $<$ 4$\mu$K), the effect of thermal velocity spread is minimized, allowing a secondary, temperature-independent mechanism to become visible. This effect stems from the spatial intensity inhomogeneity of the laser beam, which was explicitly included in our simulation. In our model, we did not assume a uniform plane wave; instead, the laser intensity was modeled with an inhomogeneous spatial distribution across the transverse plane. As a result, atoms at different positions within the finite-sized atomic cloud experience different local laser intensities and thus different Rabi frequencies. This leads to a distribution of pulse area errors across the entire atomic ensemble, which limits the final contrast even when thermal motion is negligible. The slight performance advantage of the Gaussian pulse in this low-temperature, inhomogeneity-dominated regime may suggest that its specific spatial intensity profile provides a more favorable averaging effect over the atomic cloud's distribution compared to the flatter super-Gaussian profile in the near-stationary limit.

As shown in Table \ref{table2}, different pulse sequences exhibit distinct effects on interference fringe contrast across temperature regimes. Under the cryogenic condition of 0.1~$\mu$K atomic cloud temperature, the super-Gaussian pulse sequences demonstrate superior contrast performance compared to rectangular pulses while approaching that of Gaussian pulses. Notably, the contrast exhibits asymptotic convergence with increasing pulse order, in full agreement with the results presented in figure \ref{figure6}. When the atomic cloud temperature reaches 5~$\mu$K, although all pulse sequences show reduced contrast, super-Gaussian pulses maintain relatively better performance, confirming their temperature robustness. Among all super-Gaussian orders, the 4th-order pulse achieves maximum contrast values of 0.8254 at 0.1~$\mu$K and 0.1709 at 5~$\mu$K. Based on these findings, we focus subsequent analysis on 4th-order super-Gaussian pulses to systematically evaluate their robustness and control performance under varied experimental conditions.

\begin{table}[!htbp]
\centering
\caption{Comparison of interference contrast for different pulse shapes at 0.1$\mu$K and 5$\mu$K, along with their relative enhancement percentages compared to rectangular pulses. Contrast data represent averages of 50 measurements at each temperature.\label{table2}}
\begin{tabularx}{\linewidth}{>{\centering\arraybackslash}X >{\centering\arraybackslash}X >{\centering\arraybackslash}X >{\centering\arraybackslash}X >{\centering\arraybackslash}X}
\toprule
\multirow{2}{*}{Pulse Sequence} & \multicolumn{2}{c}{Contrast} & \multicolumn{2}{c}{Improvement over Rec pulses} \\
\cmidrule(lr){2-3} \cmidrule(lr){4-5}
	& 0.1$\mu$K & 5$\mu$K & 0.1$\mu$K & 5$\mu$K \\
\midrule
Rec & 0.7733 & 0.0895 & -- & -- \\
Gaussian & 0.8181 & 0.1523 & 5.7934\% & 70.1676\% \\
SG2 & 0.8244 & 0.1684 & 6.6080\% & 88.1564\% \\
SG3 & 0.8252 & 0.1707 & 6.7115\% & 90.7263\% \\
SG4 & 0.8254 & 0.1709 & 6.7374\% & 90.9497\% \\
SG5 & 0.8253 & 0.1707 & 6.7244\% & 90.7263\% \\
SG6 & 0.8253 & 0.1704 & 6.7244\% & 90.3911\% \\
SG7 & 0.8252 & 0.1701 & 6.7115\% & 90.0559\% \\
SG8 & 0.8251 & 0.1699 & 6.6986\% & 89.8324\% \\
SG9 & 0.8251 & 0.1696 & 6.6986\% & 89.4972\% \\
SG10 & 0.8250 & 0.1694 & 6.6856\% & 89.2737\% \\
\bottomrule
\end{tabularx}
\end{table}

In the pulsed interferometry of an atom interferometer, the tolerance of pulse duration and off-resonance errors are critical factors for ensuring system stability and measurement accuracy. Precise control of the pulse duration is essential for achieving efficient Raman transitions, while off-resonance errors may lead to frequency detuning, thereby affecting the contrast of interference fringes and the sensitivity of the system. Therefore, accurately evaluating and optimizing the tolerance ranges of these errors is of great significance for enhancing the robustness and anti-interference capability of atom interferometers in complex experimental environments.

\begin{figure}[!htbp]
\begin{adjustwidth}{-0.3cm}{2cm}
\includegraphics[scale=0.50]{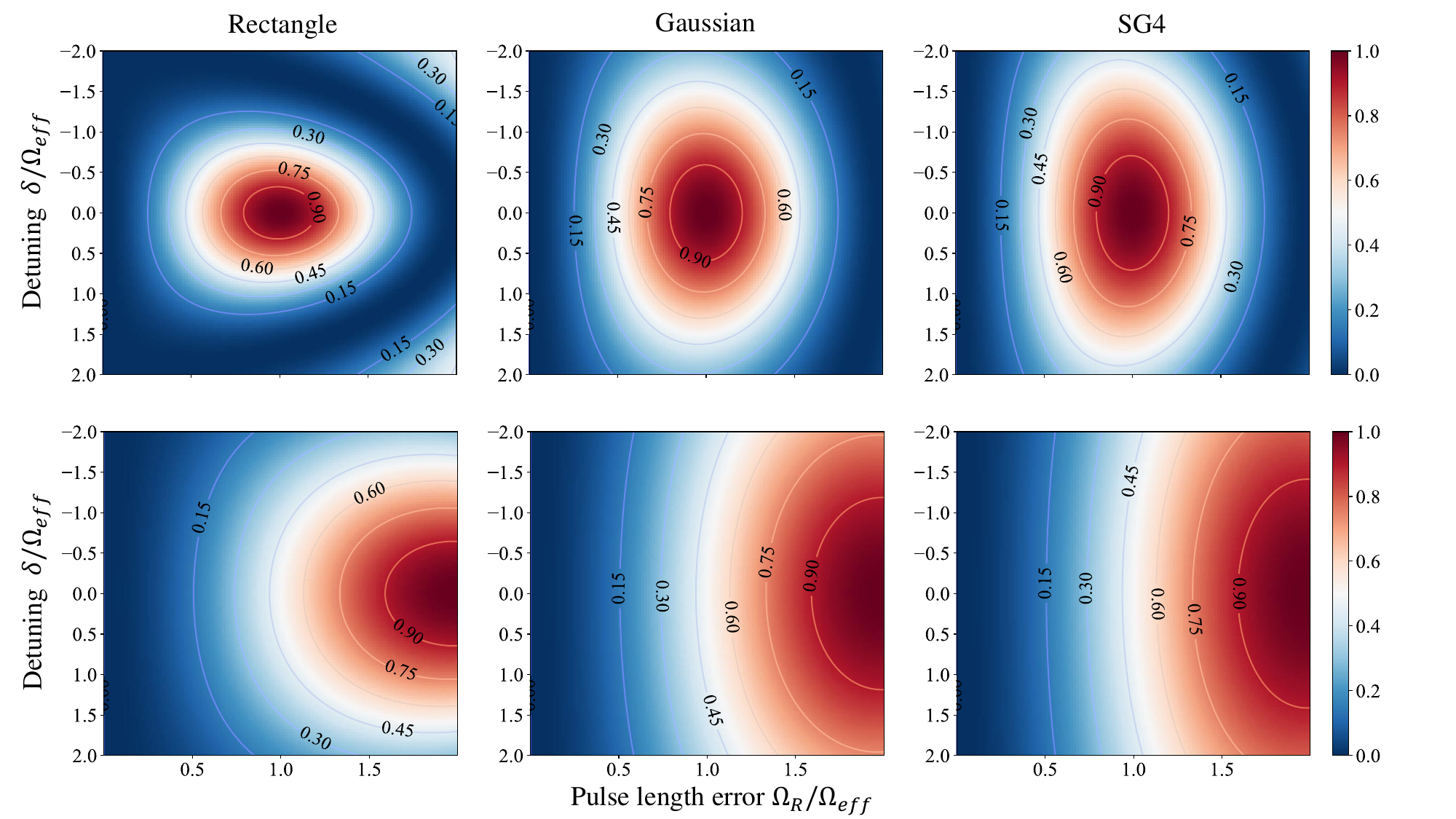}
\end{adjustwidth}
\caption{Final excited-state population distributions of different pulse profiles under detuning and pulse-length errors. Top panel: $\pi$-pulse, for which the ideal transition probability is 1.0. bottom panel: $\pi/2$-pulse, for which the ideal outcome is a 0.5 transition probability to create a perfect superposition state.\label{figure7}}
\end{figure}

Taking the pulse shapes shown in figure \ref{figure3} as examples, we performed numerical simulations to analyze the variation of interferometer contrast under different detuning magnitudes and coupling strengths (i.e., the effective Rabi frequency, $\Omega_{\text{eff}}$). figure \ref{figure7} presents a comparative study of normalized transition efficiency among rectangular pulses, Gaussian pulses, and fourth-order super-Gaussian pulse sequences, considering both pulse duration errors and finite detuning ranges. The simulation results demonstrate that Gaussian-type pulse sequences (including both Gaussian and super-Gaussian pulses) exhibit significantly enhanced robustness and stability compared with conventional rectangular pulses when facing detuning and coupling strength errors, effectively suppressing contrast degradation caused by these imperfections. Particularly noteworthy is that this robustness can be quantified by the area of the high-fidelity parameter space. As shown in figure \ref{figure7}, the 4th-order super-Gaussian pulse sequence can maintain transition fidelity above 90\% in a parameter region that is about 1.5 times larger than that of rectangular pulse sequences and, crucially, 1.1 times larger than that of Gaussian pulse sequences. This quantitatively demonstrates its superior robustness against combined pulse-length and detuning errors, which is a key advantage not apparent from comparing single-point peak contrast values alone.

\subsection{Experimental procedure}
To evaluate the adaptability of rectangular, Gaussian, and super-Gaussian pulses to Doppler-broadened velocity distributions, we employ the aforementioned method to numerically compute pulse profiles and their corresponding effective Rabi frequencies. By testing different Raman pulse shapes, we characterize the relationship between transition probability, pulse profile, and detuning. 

Our numerical simulations assume a simplified two-level Raman transition system derived from the three-level system shown in figure \ref{figure2}. The cold atomic cloud comprises $5\times10^4$ \textsuperscript{87} Rb atoms, with their thermal motion in position and velocity obeying the Gaussian distribution described earlier. It is important to note that, consistent with the conditions for which our pulses are designed, these simulations were performed without a preliminary velocity selection step. Experimentally, the cloud undergoes free-fall along the Raman beam direction under gravity while exhibiting free expansion dynamics.

The pulse configuration employs counter-propagating beams, with Raman transitions set between the hyperfine states $\ket{S^2 S_{1/2}, F=1}$ and $\ket{S^2 S_{1/2}, F=2}$. figure \ref{figure4} displays the transition probabilities for different pulse profiles. The transition probability after the pulse is calculated as follows\cite{saywell2020biselective}:
\begin{equation}
P_e = |\bra{e}\ket{\psi(\tau)}|^2 = |\bra{e}U\ket{\psi_0}|^2 \label{equation19}
\end{equation}
where $P_e$ is evaluated as a function of detuning, with the pulse discretization parameter $N=128$. The $\pi$-pulse duration $\tau_{\pi}$ is set to $10\,\mu\text{s}$, while the $\pi/2$-pulse duration $\tau_{\pi/2}$ equals $5\,\mu\text{s}$. The Rabi amplitudes are adaptively adjusted according to pulse shapes to satisfy the aforementioned pulse area conditions.

In calculating the evolution of output states during interferometer phase scanning, the final interference fringes can be acquired by scanning the phase of Raman pulses throughout the interferometric process. For pulse sequences with specific profiles, under the piecewise constant approximation method, we extend equation (\ref{equation9}) to the entire interferometric process, obtaining the total propagator expression:
\begin{equation}
U_T = \left( U_{\pi/2}^{1/N} U_{\pi/2}^{2/N} \cdots U_{\pi/2}^{m/N} \right) U_{F1} \left( U_{\pi/2}^{1/N} U_{\pi/2}^{2/N} \cdots U_{\pi/2}^{m/N} \right) U_{F2} \left( U_{\pi/2}^{1/N} U_{\pi/2}^{2/N} \cdots U_{\pi/2}^{m/N} \right) \label{equation20}
\end{equation}
where $U_{F1} = \exp(i\alpha_z \Phi_1 / 2)$ and $U_{F2} = \exp(i\alpha_z \Phi_2 / 2)$ describe the phase accumulation during the free evolution time $T_F$ in the interferometric process, respectively. This definition is consistent with the free evolution operator derived from the Hamiltonian where $\Omega_{\text{eff}} = 0$. Ultimately, at specific initial temperatures, we obtain the total output probability of the atom interferometer signal as a function of phase:
\begin{equation}
P_e = \int_v dv\, P_e(v)g(v)
\label{equation21}
\end{equation}
The interference signal contrast \( C \) at a specific atomic cloud temperature can be calculated using equation (\ref{equation21}) as \( \max(P_e) - \min(P_e) \). 

\subsection{Interferometric Performance and Robustness Analysis}

Figure \ref{figure7} provides a detailed map of the final excited-state population under combined pulse-length and detuning errors, but the criteria for "high efficiency" are fundamentally different for the two types of pulses.
For the $\pi$ pulse (top panel), the goal is a complete population inversion. Therefore, the ideal outcome corresponds to an excited-state population of 1.0, which is represented by the center of the dark red region. The plot shows that this peak performance is tightly localized around the ideal point (zero detuning and a pulse-length error factor of 1), as any deviation reduces the population transfer. For the $\pi$/2 pulse (bottom panel), the goal is to create a perfect superposition state, which corresponds to a final excited-state population of exactly 0.5. Therefore, the region of "ideal performance" is not the red area, but the specific contour where the population is 0.5 (represented by the transition between the light blue and orange/white regions in the plot). The shape of this 0.5-population contour is what reveals the pulse's robustness. The noticeable tilt of this 0.5-population contour demonstrates a powerful compensation effect between pulse-length error and detuning. This behavior is explained by the Bloch sphere dynamics. An ideal $\pi$/2 pulse rotates the state vector precisely to the equator. While a pulse-length error would normally cause the state to "over-rotate" or "under-rotate" past the equator, a corresponding non-zero detuning can be introduced to counteract this. The detuning tilts the effective rotation axis, altering the state's trajectory. The tilted 0.5 contour line represents the exact set of (pulse length error, detuning) combinations that perfectly compensate for each other, guiding the state vector to land precisely on the equator. The key finding here is that for the super-Gaussian pulse, this 0.5-population contour is broader and less curved, meaning it is easier to achieve the desired perfect superposition state even when significant experimental errors are present.

\begin{figure}[H]
\begin{adjustwidth}{3.5cm}{2cm}
\includegraphics[scale=0.38]{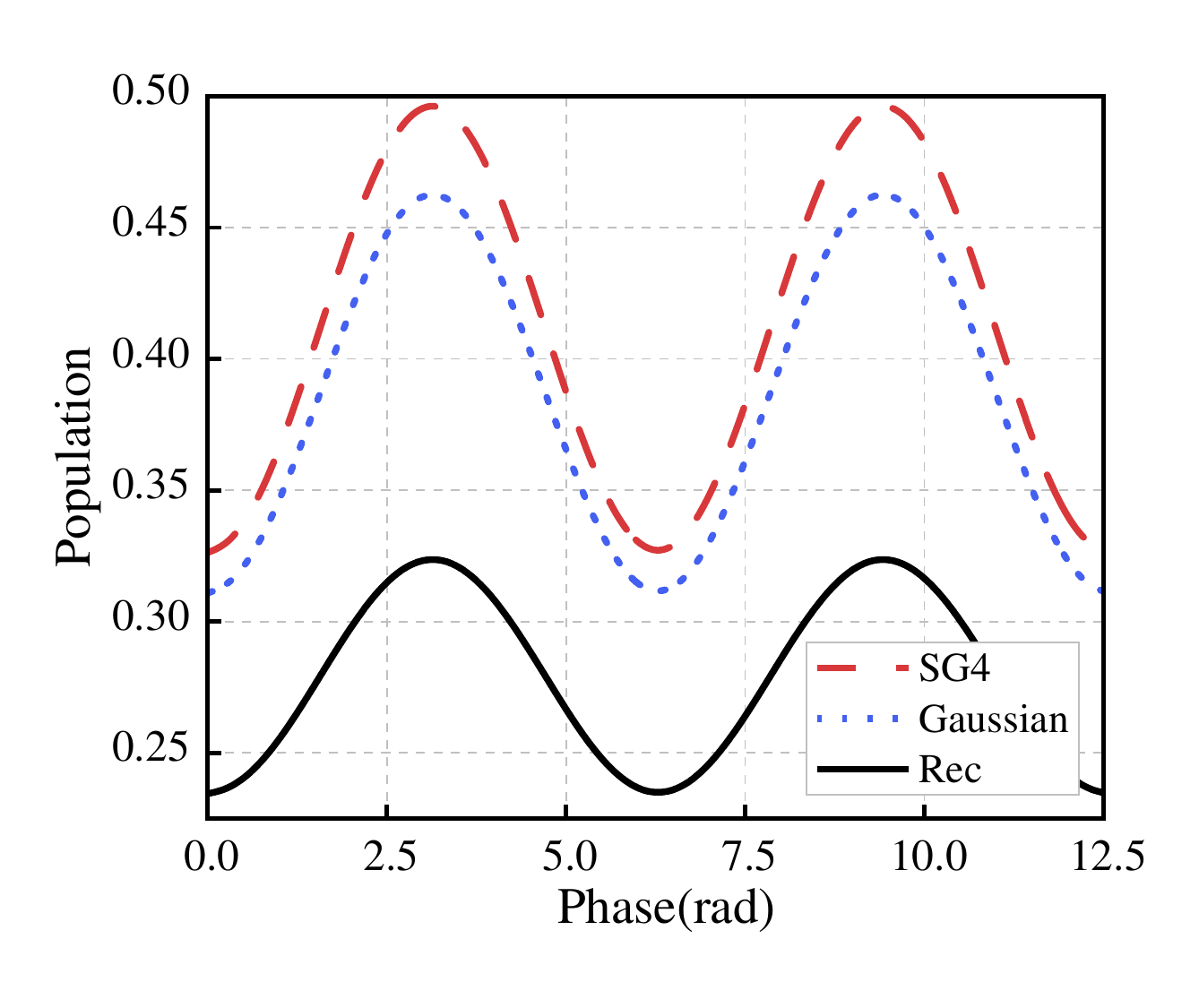}
\end{adjustwidth}
\caption{Interference fringes of rectangular, Gaussian, and 4th-order super-Gaussian pulses obtained from a 5$\mu$K atomic cloud without velocity selection. The output transition probability is shown as a function of the interferometer pulse phase. The black curve represents the reference interference fringes from rectangular pulses. The laser beam radius in these simulations is 10 mm.\label{figure8}}
\end{figure} 
As shown in figure \ref{figure8}, the initial temperature $T$ was set to 5$\mu$K, corresponding to an initial velocity of 7~mm/s for Rb atoms. The interference fringes generated by different pulse waveforms exhibit significant differences. The quantum transition probabilities of Gaussian-family pulses are overall superior to those of rectangular pulses. Notably, the super-Gaussian pulse not only achieves the highest transition probability but also significantly improves the interference fringe contrast. The 4th-order super-Gaussian pulse demonstrates a fringe contrast of 0.17, exceeding the values of 0.15 for Gaussian pulses and 0.08 for rectangular pulses, representing approximately 20\% and 90\% improvements, respectively. Compared with rectangular and Gaussian pulse sequences, the 4th-order super-Gaussian pulse sequence shows clear advantages in enhancing interference fringe contrast. The results indicate that 4th-order super-Gaussian pulses exhibit stronger robustness against laser frequency fluctuation and pulse length errors in atom interferometry, and can form $\pi/2$-$\pi$-$\pi/2$ pulse sequences suitable for high-fidelity quantum control.

\section{Conclusion}

This work systematically investigated the effects of different laser pulse shapes in MZ atom interferometers on $^{87}$Rb atomic systems, with a focus on a comparative performance analysis of rectangular, Gaussian, and super-Gaussian pulses. Our numerical simulations, based on a simplified two-level model, demonstrate that pulse shaping is a critical factor in mitigating contrast degradation, particularly for interferometers operating with thermal atomic clouds.

The results show that under thermal conditions where Doppler broadening is significant, 4th-order super-Gaussian pulses can nearly double the interference contrast compared to traditional rectangular pulses and exhibit enhanced robustness over standard Gaussian pulses. Furthermore, we identified a saturation effect where increasing the super-Gaussian order beyond n=4 yields diminishing returns, indicating a practical optimization limit. While our two-level model serves as an effective proof-of-principle, we acknowledge that a complete description would involve a three-level system, which may introduce additional effects. Nonetheless, our findings provide a valuable guide for experimentalists, demonstrating that super-Gaussian pulse shaping is a highly effective strategy for improving the performance and stability of atom interferometers that contend with significant atomic thermal motion.

Future work will focus on extending these simulations to a full three-level model to investigate the interplay between pulse shaping and the complexities of the Raman transition, as well as exploring joint optimization of pulse shapes and phases to further improve interferometer signal quality and system stability.
\section*{Acknowledgements}
This article is supported by the National Key R\&D Program (2023YFC2907002), Deep Earth National Science and Technology Major Project of China (2024ZD1003001).

\section*{Data Availability Statement}
All data that support the findings of this study are included within the article.

\bibliographystyle{iopart-num}
\bibliography{sample}
\end{document}